\documentclass[prd,nofootinbib,showpacs]{revtex4}
\usepackage{epsfig}

\newcommand{\be}{\begin{equation}}\newcommand{\ee}{\end{equation}}
\newcommand{\bea}{\begin{eqnarray}}\newcommand{\eea}{\end{eqnarray}}
\newcommand{\bit}{\begin{itemize}}\newcommand{\eit}{\end{itemize}}
\newcommand{\ben}{\begin{enumerate}}\newcommand{\een}{\end{enumerate}}

\def\lab{\label}
\def\lf{\left}

\def\non{\nonumber}\def\pa{\partial}\def\ran{\rangle}

\def\ri{\right}\def\ti{\tilde}
\def\al{\alpha}\def\bt{\beta}\def\ga{\gamma}\def\Ga{\Gamma}
\def\de{\delta}\def\De{\Delta}\def\ep{\epsilon}
\def\te{\theta}
\def\la{\lambda}\def\si{\sigma}
\def\om{\omega}\def\Om{\Omega}

\def\1{{_{1}}}\def\2{{_{2}}}
\def\bp{{\bf {p}}}\def\bk{{\bf {k}}}\def\bx{{\bf {x}}}

\def \ak{\alpha^r_{{\bf k},e}(0)}\def \akd{\alpha^{r\dag}_{{\bf k},e}(0)}

\def\twoint{\int\!\!\!\int}

\def\bbq{{\bf {q}}}\def\q{{q}}

\newcommand{\ide}{1\hspace{-1mm}{\rm I}}
\begin{document}

\title{Neutrino oscillations from relativistic flavor currents}

\author{Massimo Blasone }
\affiliation{The Blackett Laboratory, Imperial College London, London SW7
2AZ, U.K.}
\affiliation{Dipartimento di
Fisica and INFN, Universit\`a di Salerno, 84100 Salerno, Italy }
\author{Paulo Pires Pach\^eco}
\affiliation{The Blackett Laboratory, Imperial College London, London
SW7 2AZ, U.K.}
\author{Hok Wan Chan Tseung}
\affiliation{Department of Physics, University of Oxford, Oxford
OX1 3LB, U.K.}



\begin{abstract}
By resorting to recent results on the relativistic currents for
mixed (flavor) fields, we calculate a space-time dependent
neutrino oscillation formula in Quantum Field Theory. Our
formulation provides an alternative to  existing approaches for
the derivation of space dependent oscillation formulas and it also
accounts for the corrections due to the non-trivial nature of the
flavor vacuum. By exploring different limits of our formula, we
recover already known results. We study in detail the case of
one-dimensional propagation with gaussian wave-packets both in the
relativistic and in the non-relativistic regions: in the last
case, numerical evaluations of our result show   significant
deviations from the standard formula.
\end{abstract}

\maketitle


\section{Introduction}

The recent experimental evidence\cite{homestake,Kam,SNO,kamland,K2K} of
neutrino mixing and oscillations\cite{Pont} is  the first clear
sign of physics beyond the Standard Model\cite{Cheng-Li}. Thus
much effort is currently devoted to the full understanding of such
phenomenon, from the issue of its origin to more phenomenological
ones.

In this framework, there has been recently a remarkable progress
in the study of the problem of field mixing in Quantum Field
Theory (QFT)
\cite{BV95,BHV99,bosonmix,3flav,remarks,currents,comment,Berry,hannabuss,Ji,fujii1,fujii2,fujii3}.
It is important to remark that, in spite of the fact of being
usually treated in Quantum Mechanics (QM), the mixing between
states of different masses is not even allowed in a
non-relativistic theory due to the Bargmann superselection rules
\cite{Barg}. The problem of constructing an Hilbert space for
flavor states is indeed a long standing one \cite{Fujiiold} which
has even been  claimed to be impossible to achieve \cite{Kimbook}
(see however Ref.\cite{fujii1} for a criticism of that argument).
As first shown in Ref.\cite{BV95}, the difficulty lies in the fact
that the Hilbert spaces for particles with definite flavor and
those with definite mass are actually orthogonal in the infinite
volume limit. This result has  been subsequently generalized for
any number of generations and for different types (fermionic or
bosonic) of fields\cite{hannabuss,Ji,bosonmix}. It has also
emerged that the use of the correct Hilbert space (i.e. the one
for the flavor fields) leads to corrections to the usual
Pontecorvo oscillation formula.

The necessity for a full QFT treatment of mixing also stems from a
more phenomenological perspective, namely from the calculation of
(flavor) oscillation formulas. The usual treatment gives indeed in
a very simple way an oscillation formula in time, which
is actually not of great use when discussing current experiments
where the distance source-detector is measured rather than time
(data being collected over large time intervals)\cite{Kayser}. In
order to derive an oscillation formula in space, one usually
converts the time oscillation formula by means of some assumptions
(equal time assumption, classical propagation, etc.) which are
however questionable in many respects (see Ref.\cite{Beuthe} for a
review). Thus several QFT approaches have been developed
\cite{giuntipak,Grimus,Cardall,Kiers,Ishikawa}, such as the {\em
external wavepacket models} or the {\em stationary boundary
conditions models}, which attempt  a more realistic
description of neutrino oscillations.
It seems indeed that the main features of such
formula are now quite well understood \cite{Beuthe}, including the
concepts of coherence length, localization terms, dispersion times, etc.

On the other hand, all these calculation schemes have been
developed for highly relativistic neutrinos and/or for nearly degenerate
masses and in most cases the spin structure of the wavefunctions was
neglected (i.e. neutrinos were treated as scalars).
Also, the identification of the Hilbert spaces for mass and flavor neutrinos
was implicitly assumed in the choice of the propagators (see Ref.\cite{BHV99} for
a discussion of flavor propagators).

A completely different approach to the problem is indeed
possible and was advocated in Refs.\cite{Anco,Zralek}:
an oscillation formula in space could
be derived in a straightforward way,
by considering the flux of neutrinos through the detector as the integral over the
measurement time and detector surface of the
relativistic flavor current for the oscillating neutrinos.
An attempt in this direction is contained in Ref.\cite{Anco},
which however deals with QM and therefore
uses a probability current derived from Schr\"odinger equation.
As pointed out in Ref.\cite{Zralek}, the
obstacle to the extension of this result to
QFT lies in the difficulty of constructing a
relativistic current for mixed particles, which boils
down again to the problem of the definition of
the Hilbert space for such (not on-shell) particles.

The solution to this problem was recently given in
Ref.\cite{currents}: it turned out that it is indeed
possible to define these (non-conserved) flavor currents
in a consistent way by use of previous results on
 the flavor Hilbert space. An analysis of the
currents associated to mixing for the case
of three flavor neutrino mixing with CP violation and for boson mixing can be
found in Refs.\cite{3flav,bosonmix}.

In this paper, we calculate the oscillation
formula by using the relativistic flavor currents above
mentioned. For simplicity, we only treat the case of mixing
among two generations of Dirac neutrinos: the extension to Majorana fields and to
three flavors will be given elsewhere. The formula we obtain in a very
direct way has the full space-time dependence and contains previous results
obtained by use of the flavor charges. We then obtain a general expression for the
electron neutrino flux in three dimensions with spherical symmetry.
Although a fully three dimensional analysis is possible within
our formulation, we study in detail the one-dimensional case with gaussian wavepackets,
which is indeed the one most frequently considered in literature.
We show how in the highly relativistic limit and by integrating over an infinite time,
the standard space dependent oscillation formula\cite{giuntipak,Beuthe} is obtained,
which includes the coherence length and the localization term.
However our formula is also valid for  non-relativistic neutrinos and it accounts
for flavor vacuum effects. Numerical evaluations of our result show
deviations from the usual formula in the non-relativistic regime.

The paper is organized as follows: in Section II we present
a derivation of the flavor currents in the
line of Ref.\cite{currents} and give some details on
the construction of the flavor Hilbert space.
In Section III we first obtain the expectation value of the flavor current
in the most general case and then  a more explicit
expression in the case of a spherically symmetric emission.
Then, in Section IV, we specialize to the one-dimensional
case and obtain the space dependent expression for the
electron neutrino flux.  Further analysis is done by choosing
gaussian wavepackets.
Section V is devoted to conclusions.

\section{Relativistic currents for mixed fields}

Following Ref.\cite{currents}, let us consider the
following Lagrangian density describing two Dirac fields  with a
mixed mass term:
\bea\label{lagemu} {\cal L}(x)\,=\,  {\bar \Psi_f}(x) \lf( i
\not\!\partial -
  M \ri) \Psi_f(x)\, ,
\eea
where $\Psi_f^T=(\nu_e,\nu_\mu)$ and $M = \lf(\begin{array}{cc} m_e &
m_{e\mu} \\ m_{e\mu} & m_\mu \end{array} \ri)$.
The mixing transformations
\bea\label{fermix}
\Psi_f(x) \, =\, \lf(\begin{array}{cc}  \cos \te & \sin \te \\
-\sin \te & \cos \te \end{array}\ri) \Psi_m (x) \, , \eea
with $\te$ being the mixing angle and $\Psi_m^T=(\nu_1,\nu_2)$,
diagonalize the quadratic form of Eq.(\ref{lagemu}) to the
Lagrangian for two free Dirac fields, with masses $m_1$ and $m_2$:
\bea\label{lag12} {\cal L}(x)\,=\,  {\bar \Psi_m}(x) \lf( i
\not\!\partial -
  M_d\ri) \Psi_m(x)  \, ,
\eea
where $M_d = diag(m_1,m_2)$. One also has $ m_{e} =
m_{1}\cos^{2}\te + m_{2} \sin^{2}\te~$,  $m_{\mu} =
m_{1}\sin^{2}\te + m_{2} \cos^{2}\te~$,   $m_{e\mu}
=(m_{2}-m_{1})\sin\te \cos\te\,$. We assume $m_2\ge m_1$
and $\te$  in $[0,\frac{\pi}{4}]$.

The Lagrangian Eq.(\ref{lagemu}) is invariant under global
$U(1)$ phase transformations  of the type $\Psi_m' \, =\, e^{i \al
}\, \Psi_m$: as a result, we have the conservation of the Noether
charge $Q=\int d^3\bx \, I^0(x) $ (with $I^\mu(x)={\bar \Psi}_m(x)
\, \ga^\mu \, \Psi_m(x)$)  which is indeed the total charge of the
system (i.e. the total lepton number).

Consider now the $SU(2)$ transformations acting on
$\Psi_m$:
\bea \label{masssu2} \Psi_m'(x) \, =\, e^{i \al_j  \tau_j}\,
\Psi_m (x) \, \qquad, \qquad
 j=1, 2, 3,
\eea
with $\al_j$ real constants, $\tau_j=\si_j/2$ and $\sigma_j$ being
the Pauli matrices.

For $m_1\neq m_2$, the Lagrangian is not generally invariant under
(\ref{masssu2}) and  we obtain, by use of the equations of motion,
\bea \non &&\de {\cal L}(x)\,= \,  i \al_j \,{\bar \Psi_m}(x)\,
[\tau_j,M_d ]\, \Psi_m(x) \, =\,  - \al_j \,\pa_\mu J_{m,j}^\mu
(x),
\\ [3mm]\label{fermacu1}
&&J^\mu_{m,j}(x)\, =\,  {\bar \Psi_m}(x)\, \ga^\mu\, \tau_j\,
\Psi_m(x) \qquad, \qquad j=1, 2, 3. \eea

The above analysis is valid at classical level. We now quantize the
free fields $\nu_1$ and $\nu_2$ in the usual way
(see Appendix A for conventions).

Then the charge operators  $Q_{m,j}(t)\equiv \int d^3 \bx \,J^0_{m,j}(x) $,
satisfy the  $su(2)$  algebra $[Q_{m,j}(t), Q_{m,k}(t)]\, =\, i
\,\ep_{jkl} \,Q_{m,l}(t) $. The Casimir operator is proportional
to the total charge. From (\ref{fermacu1}) we also see  that
$Q_{m,3}$ is conserved as $M_d$ is diagonal. Let us define the
combinations:
\bea\label{noether1}
&&Q_{1}\, \equiv \,\frac{1}{2}Q \,+ \,Q_{m,3}
\quad, \quad Q_{2}\, \equiv \,\frac{1}{2}Q \,- \,Q_{m,3}
\\
&&Q_i \, = \,\sum_{r} \int d^3 \bk \lf(
\al^{r\dag}_{{\bf k},i} \al^{r}_{{\bf k},i}\, -\,
\bt^{r\dag}_{-{\bf k},i}\bt^{r}_{-{\bf k},i}\ri) \quad, \quad i=1,
2, \eea
where the last expression has been normal ordered, as usual.
These are nothing but  the Noether charges associated with the
non-interacting fields $\nu_1$  and $\nu_2$: in the absence of
mixing, they are the flavor charges,  separately
conserved for each generation.
Observe now that the transformation
\bea \Psi_f(x) &=& e^{-2 i  \te Q_{m,2}(t) } \Psi_m(x) e^{2 i \te
Q_{m,2}(t) } \eea
is just the mixing Eq.(\ref{fermix}). Thus $G_\te(t) \equiv e^{2 i
\te Q_{m,2}(t)}$ is the generator of the mixing transformations (see Appendix A).
In Ref.\cite{BV95}, it has
been shown that the action  of $G_\te(t)$ on the vacuum state $|0\ran_{1,2}$
results in a new vector (flavor vacuum) $|0(t)\ran_{e,\mu}\equiv
G^{-1}_\te(t)|0\ran_{1,2}$, orthogonal to $|0\ran_{1,2}$ in the
infinite volume limit.

Following Ref.\cite{currents}, in accordance with Eq.(\ref{noether1}),
we define the {\em flavor charges} for mixed fields as
\bea
Q_\si(t) & \equiv & G^{-1}_\te(t) \, Q_i \, G_\te(t)
\eea
with $(\si,i)=(e,1) , (\mu,2)$ and $Q_e(t)  + Q_\mu(t) =  Q$.
They have a simple expression in terms of the flavor
ladder operators:
\bea Q_\si(t) & = & \int d^3 \bx  \;\nu^\dag_\si(x) \,\nu_\si(x)\, =\,
\sum_{r} \int d^3 \bk \, \lf(
\al^{r\dag}_{{\bf k},\si}(t) \al^{r}_{{\bf k},\si}(t)\, -\,
\bt^{r\dag}_{-{\bf k},\si}(t)\bt^{r}_{-{\bf k},\si}(t)\ri)\,, \quad \si= e,\mu.
\eea
These charge operators then act
on the flavor Hilbert space ${\cal H}_{e,\mu}$  (see Appendix
A). For the single neutrino
and antineutrino states of definite momentum and helicity\footnote{Note
that there is no ambiguity  here in defining the flavor neutrino states with
definite momentum.}:
\bea\label{state}
|\nu_{\sigma}(t)\ran \equiv\;
\al^{r\dag}_{{\bf k},\si}(t)\;|0(t)\ran_{e,\mu}\quad ; \quad
|{\bar \nu}_{\sigma}(t)\ran \equiv\;\bt^{r\dag}_{{\bf
k},\si}(t)\;|0(t)\ran_{e,\mu}\, ,
\eea
one obtains
\bea
Q_\si(t)|\nu_{\sigma}(t)\ran =  |\nu_{\sigma}(t)\ran \quad ; \quad
 Q_\si(t)|{\bar \nu}_{\sigma}(t)\ran = - |{\bar \nu}_{\sigma}(t)\ran\,.
 \eea
In the following, we will use the flavor currents:
\bea
J^\mu_\si(x) & \equiv &  {\bar \nu}_\si(x) \,\ga^\mu\,\nu_\si(x)\, =\,
G^{-1}_\te(t) \, J^\mu_i(x) \, G_\te(t)\,,
\eea
with $(\si,i)=(e,1),(\mu,2)$.
From the global $U(1)$ invariance follows  the continuity equation:
\bea
\pa_\mu \lf[J^\mu_e(x) + J^\mu_\mu(x) \ri] \,=\,0\, .
\eea

\section{space-time dependent neutrino oscillation formula}

In this Section, we
consider the calculation of the general (i.e. space-time dependent)
neutrino oscillation formula, by
use of the flavor currents introduced in \S II. In Section IV,
we then extract  from this
the space dependent formula by integrating over time.

As already discussed in Ref.\cite{Anco,Zralek}, although in the context of
QM, what one is ultimately interested in current experiments
is actually the flux of
 neutrinos of a given flavor through the  surface of the detector $\Om$ in a (large)
measurement time $T$. In the case of electron neutrinos, this quantity is given by:
\bea\label{flux}
&&\Phi_{\nu_e\to \nu_e}(L)=\int_{t_0}^{T}
d t\int_{\Omega} \langle \nu_e| J_e^{i}(\bx,t)|
\nu_e\rangle\,\, d {\bf S}^i\, ,
\eea
where $L$ is the distance source-detector.
Our aim is now to calculate the expectation value of the flavor four-current density
on a localized neutrino state, defined  as a wave-packet in flavor Hilbert space.

\subsection{Expectation value of the flavor current density}

We define an initial\footnote{We work in the Heisenberg picture,
the time evolution is thus borne by the operators.}, localized
state $| \nu_e (\bx_0,t_0) \rangle$ with definite flavor {\it
e}\, :
\bea\label{initial_state}
| \nu_e(\bx_0,t_0)\rangle=A\int d^3
\bk \,\,  {e}^{-i(\om_{k,1}t_0-\bk.\bx_0)}
f(\bk)\,\,\alpha_{\bk,e}^{r\dag}(t_0) \,\,|0\rangle_{e,\mu}\, ,
\eea
where {\it A} is a normalization constant, $f(\bk)$ is the form of
the wave-packet and $|0 \rangle_{e,\mu}$ the flavor vacuum at time
$t=t_0$. This corresponds to an electron neutrino being emitted
 at $(t_0,\bx_0)$. For convenience we set $(t_0,\bx_0) = (0,0,0,0) $ in the following.
Note that the wave packet is defined in the Hilbert space of flavor fields.

The normalization of this state is  given by:
\bea\label{modanti}
\non 1&=&\langle
 \nu_e| \nu_e\rangle \,=\, |A|^2\,\int d^3 \bk \int d^3 \bp
 \,f^*(\bk)\,f(\bp)\,\{\ak,\alpha_{\bp,e}^{r\dag}(0) \}\, =\,|A|^2\,
 \int d^3 \bk\,\,|f(\bk)|^2\, .
  \eea
From Section II, the  electron neutrino four-current is given as
\bea \label{sigmacurrent}
J_e^{\,{\mu}}({\bf x},t) &=&
\bar\nu_e({\bf x},t)\gamma^{\mu}\nu_e({\bf x},t) \,=\,\nu_e^{\dag}(\bx,t) \,
\,\Gamma^{\mu} \,\,\nu_e({\bf x},t)\, ,
\eea
where $ \Gamma^{\mu}= \gamma^0\gamma^{\mu} $.
In the following we will use the chiral representation  of
the Dirac matrices (see Appendix B).
The expansion  for the flavor fields given in Eq.(\ref{flavfields})
leads to the following explicit form of the current operator:
\bea\non J_e^{\mu}(\bx,t)&=&\twoint\frac{d^3
\bp}{(2\pi)^\frac{3}{2}}\,\frac{d^3 \bk}{(2\pi)^\frac{3}{2}}\,e^{i(\bk -\bp) \cdot \bx}
\sum\limits_{r,s}\Big[
\,u^{s\dag}_{\bp,1}(t)\,\Gamma^{\mu}\,u_{\bk,1}^r(t)\,
\alpha_{\bp,e}^{s\dag}(t)\alpha_{\bk,e}^r(t)-
v^{s\dag}_{-\bp,1}(t)\,\Gamma^{\mu}\,v_{-\bk,1}^r(t)\,
\beta_{-\bk,e}^{r\dag}(t)\beta_{-\bp,e}^s(t)
\\ \label{current}
&+& u^{s\dag}_{\bp,1}(t)\,\Gamma^{\mu}\,v_{-\bk,1}^r(t)
\,\alpha_{\bp,e}^{s\dag}(t)\beta_{-\bk,e}^{r\dag} (t)+
v^{s\dag}_{-\bp,1}(t)\,\Gamma^{\mu}\,u_{\bk,1}^r(t)
\,\beta_{-\bp,e}^{s}(t)\alpha_{\bk,e}^r(t)\, \Big]\, .
\eea
In Appendix \ref{appenvac}, we prove that
$_{e,\mu}\langle0|J^{\mu}(\bx,t)|0\rangle_{e,\mu}=0$. This
result leads to (see Appendix A):
\bea\label{J1}
&&\langle \nu_e |J_e^\mu(\bx,t)| \nu_e\rangle
\,=\, \Psi^\dag(\bx,t)\, \Ga^\mu \lf(\begin{array}{cc}1&1\\1&1\end{array}\ri)
\,\Psi(\bx,t)
\eea
with
\bea
&& \Psi(\bx,t) \equiv A \int\frac{d^3 \bk}{(2\pi)^\frac{3}{2}}\,
e^{i\bk \cdot \bx}\,f(\bk)\,
\lf(\begin{array}{l} u_{\bk,1}^r \, X_{\bk,e}(t)
\\ [3mm]
 \sum_{s} v_{-\bk,1}^s \,(\vec{\si}\cdot \bk)^{sr}\, Y_{\bk,e}(t)
 \end{array} \ri)
\\ [2mm]
&&X_{\bk,e}(t)\,=\,\cos^2\theta e^{-i\om_{k,1}t}+\sin^2\theta
\left[e^{-i\om_{k,2} t}|U_{\bk}|^2+
e^{i \om_{k,2}t}|V_{\bk}|^2\right]
\,
\\[2mm]
&&Y_{\bk,e}(t) \,=\,\sin^2\theta|U_{\bk}|\chi_1\chi_2
\left[\frac{1}{\om_{k,2}+m_2}  -
\frac{1}{\om_{k,1}+m_1}\right]\left[e^{-i \om_{k,2} t}-
e^{i \om_{k,2}t}\right]
\eea
where  $\vec{\sigma} \cdot \bk  = \lf(\begin{array}{cc} k_3 & k_-\\ k_+ & -k_3\end{array}\ri)$
and $\chi\,_{i}\equiv
\left(\frac{\omega_{k,i}+m_{i}}{4 \omega_{k,i}}\right)^{\frac{1}{2}}$.
We made use of the following relations:
\bea\label{anticommutators}
&&\{\alpha_{\bk,e}^r(t),\,\al_{\bp,e}^{s\dag}(0)\}=
\de^{rs} X_{\bk,e}(t)e^{i \om_{k,1} t}
\,\de^3(\bk-\bp)
\quad, \quad
\{\bt_{-\bk,e}^{r\dag}(t),\,\al_{\bp,e}^{s\dag}(0)\}
=  (\vec{ \si}\cdot \bk)^{rs}\,
Y_{\bk,e}(t) e^{-i \om_{k,1} t}\, \de^3(\bk-\bp)\, .
\eea

The expression in Eq.(\ref{J1}) contains the most
general information about neutrino oscillations and
can explicitly evaluated  once
the form of the wave-packet is specified. A similar expression can be easily obtained for
the other quantity of interest, namely $\langle \nu_e
|J_\mu^{\,{\mu}}(\bx,t)| \nu_e\rangle$.

For comparison with previous results and for better understanding
the expression (\ref{J1}), let us consider the limit situation in
which the wave-packet becomes a plane wave, with definite
momentum: $f(\bk)= (2\pi)^\frac{3}{2} \delta^3(\bk-\bbq)$.
This obviously means that we loose information about localization
and we need to integrate over the entire volume to get the flavor
oscillations (in time). We indeed find ($\rho_e\equiv J_e^0$):
\bea
\langle \nu_e|\,Q_e(t)\,| \nu_e\rangle\,=\,
\int d^3\bx\, \langle \nu_e|\,\rho_e(x)\,| \nu_e\rangle &= &
1-\sin^2(2\theta)
 \left[|U_{\bbq}|^2\sin^2\left(\frac{\om_{\q,2}-\om_{\q,1}}{2}t\right)+|V_{\bbq}|^2\sin^2
 \left(\frac{\om_{\q,2}+\om_{\q,1}}{2}t\right)\right]
 \eea
which agrees with the result obtained in Refs.\cite{BHV99}. Notice
the presence of the non-standard oscillation term and of the
momentum dependent amplitudes (the Bogoliubov coefficients
satisfying $|U_{\bbq}|^2+|V_{\bbq}|^2=1$).

We now consider the situation in which we have a
spherically symmetric emission of the
neutrinos from the source at $(x_0,t_0)$. This allows us to limit our
investigation to the radial flux, which can be identified without loss of
generality with the $z$- component of the current. In this case,
Eq.(\ref{J1}) takes it simplest form, since $\Gamma^3$ is diagonal in
the chosen (chiral) representation.

The  matrix $\Gamma^3$  can be expanded
as a linear combination of  spinor outer products:
\bea \label{sigma3}
\Gamma^3=\left(\begin{array}{cccc}1&0&0&0\\ 0&-1&0&0\\ 0&0&-1&0
\\ 0&0&0&1\end{array}\right)=\gamma^0\gamma^3=
\sum\limits_{j=1}^4\,(-1)^{j +1}\,\eta_j\eta_j^{\dag}\,
\eea
where
\bea
 \eta_1=\left(\begin{array}{c}1\\0\\0\\0\end{array}\right),\quad\eta_2=
\left(\begin{array}{c}0\\1\\0\\0\end{array}\right),\quad\eta_3=
\left(\begin{array}{c}0\\0\\0\\1\end{array}\right),\quad\eta_4=
\left(\begin{array}{c}0\\0\\1\\0\end{array}\right)
\eea
This decomposition permits a remarkable rearrangement, since
Eq.(\ref{J1}) then  becomes:
\bea\label{jz}
\langle \nu_e |J_e ^3({\bx},t)|
\nu_e\rangle&=&\, \sum\limits_{j=1}^4(-1)^{j+1}\,
\Psi^\dag_j(\bx,t)\, \lf(\begin{array}{cc}1&1\\1&1\end{array}\ri)
\,\Psi_j(\bx,t)
\,=\,\sum\limits_{j=1}^4(-1)^{j+1}\, \lf|{\cal A}^r_j+{\cal B}^r_j\ri|^2
\eea
where (see also Appendix \ref{explicit})
\bea
&&\Psi_j(\bx,t)\, =\, \eta_j^\dag \Psi(\bx,t) \, \equiv\,
\lf(\begin{array}{c}{\cal A}^r_j({\bx},t)
\\[2mm]
{\cal B}^r_j({\bx},t) \end{array}\ri)\,.
\eea

Eq.(\ref{jz}) gives a compact expression for the  flavor current  expectation value
in the case of a spherically symmetric emission. It can indeed be integrated
to give a useful expression for the
electron neutrino flux at a given distance from the source. Although this
calculation is of interest, because it contains information on the three-dimensional nature
of the propagation, it will be not developed further here, and in the following Section
we will consider the case of one-dimensional propagation.
Note that in the present case (spherical symmetry), the expectation values of $J^1_e$ and
$J^2_e$ at any point can be obtained from the above expectation value of $J^3_e$.
In the general case Eq.(\ref{J1}), however, the
expectation values of $J^1_e$ and $J^2_e$ have to be
calculated independently and this can be done by use of the
decomposition of $\Ga^1$ and $\Ga^2$ given in Appendix \ref{chiral}.

\vspace{.3cm}

It remains to consider  $\rho_e$, which is the relevant measurable quantity for
experiments without angular resolution.
It is indeed easy to get its expectation value:
\bea\label{rho}
\langle \nu_e |\rho_e({\bx},t)|
\nu_e\rangle&=&\,
\sum\limits_{j=1}^4\, \lf|{\cal A}^r_j+{\cal B}^r_j\ri|^2\,,
\eea
which is related to the expectation value of $J^3_e$ as follows:
\bea\label{correction}
\langle \nu_e|J_e^3(\bx,t)|\,
 \nu_e\rangle&=&\langle \nu_e|\,\rho(\bx,t)\,|
 \nu_e\rangle\,- 2\, |{\cal A}^r_2+{\cal B}^r_2|^2\,-\,2
|{\cal A}^r_4+{\cal B}^r_4|^2 \eea
This is a significant result in the sense that it
directly contradicts the usual $x = t $ approach: if the neutrino really travelled
on a straight line then,  for points for which the Cartesian z-direction coincides
with the polar radial direction (as defined with respect to the source),
one should have\footnote{the result would be the same up to a constant if
one replaced the velocity of light by the (constant) average velocity of the neutrinos.}:
$\non\langle \nu_e|J_e^3 (x)|\, \nu_e\rangle = \langle \nu_e|\,
\rho(x)| \nu_e\rangle$.

\subsection{One--dimensional case}

Given the complexity of the three-dimensional analysis, we now
limit the problem to one spatial dimension (as do  most of other treatments
in the literature).

The wave packet now takes the form $f(\bk) = (2\pi)^\frac{3}{2} \delta(k_1)\,
\delta(k_2)\,f(k_3)$,
which corresponds to a packet with definite momentum in the $x$ - and $y$ -
direction and a momentum distribution $f(k_3)$ along the $z$ - direction.
The uncertainty principle then dictates that the corresponding spatial distribution
is that of a beam of infinite radius and uniform surface density moving in the
z-direction (a neutrino sheet).

In the one-dimensional case, we have (see Appendix \ref{explicit}): $ {\cal A}^r_2={\cal A}^r_3=0$
and ${\cal B}^r_2={\cal B}^r_3=0$ for all $ r$. In order to perform explicit evaluations, we
choose in the following $r= 1$:
this does not imply any loss of generality since the final result
is independent of $r$. We also  put ${\cal A}^1_1+{\cal B}^1_1= \psi_{+} + \psi_{+}'$
and ${\cal A}^1_4+{\cal B}^1_4 \,=\, \psi_{-} + \psi_{-}'$.

Eq.(\ref{jz}) then becomes:
\bea \label{1dflux}
\langle \nu_e |J_e^3(z,t)| \nu_e\rangle&=&
\,\lf|\,\psi_{+}(z,t) + \psi_{+}'(z,t)\,\ri|^2
\,-\,\lf|\,\psi_{-}(z,t) + \psi_{-}'(z,t)\,\ri|^2
\eea
with
\bea \label{general1} \psi_{+}(z,t)
&=&\frac{A}{2}\int_0^\infty \!dk\,
\Big[\cos^2\theta\,{e}^{-i\om_{k,1}t}+\sin^2\theta\,
{e}^{-i\om_{k,2}t}\Big]  {\mathrm T}_k(z)
\\\label{general2} \psi_{+}'(z,t)\,&=&iA\int_0^\infty\!{d}k\,
\sin^2\theta\,\sin(\om_{k,2}t)|V_{k}|\, \Big[ {\mathrm T}_k(z)|V_{k}| -
 {\mathrm T}_{-k}(z)|U_{k}|
\Big] \\
 \label{general3} \psi_{-}(z,t)
&=&\frac{A}{2}\int_0^\infty\!{d}k\,
\Big[\cos^2\theta\,{e}^{-i\om_{k,1}t}+\sin^2\theta\,
{e}^{-i\om_{k,2}t}\Big]  {\mathrm T}_{-k}(z)
\\\label{general4} \psi_{-}'(z,t)&=& iA\int_0^\infty\!{d}k\,
\sin^2\theta\,\sin(\om_{k,2}t) |V_{k}|\,
\Big[ {\mathrm T}_{-k}(z)|V_{k}| +  {\mathrm T}_k(z)|U_{k}|
\Big] \eea
where
\bea
 {\mathrm T}_k(z) = \,{e}^{ikz}\,f(k)\,
\lf[\lf(1 +\frac{m_1}{\om_{k,1}}\ri)^{\frac{1}{2}} \,+\,
\lf(1 -\frac{m_1}{\om_{k,1}}\ri)^{\frac{1}{2}}
\ri]
\,+\, {e}^{-ik
z}\,f(-k)\,\lf[\lf(1 +\frac{m_1}{\om_{k,1}}\ri)^{\frac{1}{2}} \,-\,
\lf(1 -\frac{m_1}{\om_{k,1}}\ri)^{\frac{1}{2}}
\ri]\eea

The above relations are the main result of this work: starting with an arbitrary
wave-packet $f(k)$ we can now determine the expected one--dimensional flux of
electron neutrinos at a given distance from the source as:
\bea\label{flux1D}
&&\Phi_{\nu_e\to \nu_e}(z)=\int_{0}^{\infty}
d t\,  \langle \nu_e| J_e^{3}(z,t)| \nu_e\rangle\,
\eea

Note that the  $\psi'_\pm$ terms are the corrections
 due to the use of the flavor Hilbert space in the definition of neutrino
state. Eq.(\ref{flux1D}) is an exact expression for the space
oscillation formula in 1D and will be in the following evaluated
numerically with the choice of a gaussian wave packet. In the
following Section, however, we digress on the relativistic limit,
which allows us to extract an (approximated) analytical  form of
Eq.(\ref{flux1D}) and to compare it with the existing results in
the literature.

\section{Approximations and Numerical Analysis}

In this Section, we explicitly evaluate
Eq.(\ref{flux1D}) by choosing a gaussian wave packet. We
first consider the relativistic limit and
show how the full result Eq.(\ref{flux1D}) reduces to the standard
formula given in \cite{giuntipak,Beuthe}.
Then we proceed with numerical evaluations of
the exact formula Eq.(\ref{flux1D}), which allow
us to test the non-relativistic regime.

We choose $f(k)$ to be  a gaussian wave-packet with  momentum
spread $\sigma_{k}$:
\bea\label{gaussian}
f(k)=\,H\,\exp\left[-\frac{(k-\q)^2}{4\sigma^2_{k}}\right] \eea
with $H=(2 \pi \si_k^2)^{-\frac{1}{4}}$.

\subsection{Recovering the standard oscillation formula: sharp gaussian wave--packets in
relativistic limit}

To recover previous results in the
literature\cite{giuntipak,Beuthe},  we consider the case of sharp
gaussian wave--packet in relativistic limit, satisfying the
following conditions :
\begin{enumerate}
\item  The mean
momentum $\q$ of both mass species lies in the relativistic regime
such that $\om_{\q,i}\gg m_i$ and $|V_{\q}|\to 0$. Hence  $\psi_{+}'$
and $\psi_{-}'$ can be ignored.

\item The gaussian is sharply
peaked, $\q\gg\sigma_{k}$: the largest contributions to the
integrals come from around the mean momentum $\q$, i.e. Laplace's
method \cite{laplace} can be used.

\end{enumerate}

Using the above  conditions, one has $\left(1\pm\frac{m_i}{\om_{k,i}}\right)^{\frac{1}{2}}
\approx 1\pm \frac{m_i}{2\om_{\q,i}}$. Thus,
to first order in
$\frac{m_1}{\om_{\q,1}}$:
\bea\non \psi_{+}(z,t)&\approx& A H\,\int_0^\infty {d}k\,\,
\left[\cos^2\te\, {e}^{-i(\om_{k,1}t-kz)}\,+\,\sin^2\te \,
{e}^{-i(\om_{k,2}t-kz)}\right]\,\exp\left[-\frac{(k-\q)^2}{4\sigma^2_{k}}\right]
\\
&+& A H \frac{m_1}{2\om_{\q,1}}\,\int_0^\infty{d}k\,
\left[\cos^2\te\, {e}^{-i(\om_{k,1}t+kz)}\,+\,\sin^2\te \,
{e}^{-i(\om_{k,2}t+kz)}\right]\,\exp\left[-\frac{(k+\q)^2}{4\sigma^2_{k}}\right]
\eea
For $\q$ positive, the integral on the  second line can be neglected.
On the other hand, we obtain for
$\psi_-$ the following expression:
\bea\non \psi_{-}(z,t)&\approx& A H
\frac{m_1}{2\om_{\q,1}}\,\int_0^\infty {d}k\,\, \left[\cos^2\te\,
{e}^{-i(\om_{k,1}t-kz)}\,+\,\sin^2\te \,
{e}^{-i(\om_{k,2}t-kz)}\right]\,\exp\left[-\frac{(k-\q)^2}{4\sigma^2_{k}}\right]
\\
&+& A H \int_0^\infty{d}k\, \left[\cos^2\te\,
{e}^{-i(\om_{k,1}t+kz)}\,+\,\sin^2\te \,
{e}^{-i(\om_{k,2}t+kz)}\right]\,\exp\left[-\frac{(k+\q)^2}{4\sigma^2_{k}}\right]
\eea
Again, the second term is exponentially suppressed and can be neglected.
The first term of $\psi_-$
is also small for relativistic neutrinos and will be neglected in the following. We thus have:
\bea \label{1dflux2}
\langle \nu_e |J_e^3(z,t)| \nu_e\rangle&\simeq&
\,|\,\psi_{+}(z,t) \,|^2
\eea

\subsubsection{Momentum integration}

We now turn towards the evaluation of $\psi_{+}$ using Laplace's
method. Let us first define the {\it dispersion}\, times
$T^{disp}_i=\om_{\q,i}/2\sigma_{k}^2$, which set the timescale for
the spatial dispersion of the wave-packets.

Expanding the integrand about the mean momentum to second order in
$\de k=k-\q$, we obtain\footnote{Using the expansion $\om_{k,i}
\approx \om_{\q,i}+\frac{\delta k^2}{2\om_{\q,i}}+v_{\q,i} \delta k $
and making a change of variable $k-\q=\delta k$, we get
$$ \psi_{+}\approx A H \,\cos^2\theta e^{i(\q z-\om_{\q,1}t)}\,
\int_{-\q}^\infty d(\delta k)\,
e^{i\delta k(z-v_{\q,1}t)}\exp \left[-\delta
k^2\left(\frac{1}{4\sigma^2_{k}}+ i
\frac{t}{2\om_{\q,1}}\right)\right]
$$
$$
+   A H \,\sin^2\theta
e^{i(\q z-\om_{\q,2}t)}\, \int_{-\q}^\infty d(\delta k)\, e^{i\delta
k(z-v_{\q,2}t)}\exp \Big[-\delta k^2\Big(\frac{1}{4\sigma^2_{k}}+
i \frac{t}{2\om_{\q,2}}\Big)\Big]
$$
which is brought in the given form by completing the square for
the argument of the gaussian.}:
\bea\non \psi_{+}(z,t)& \approx & A H\, \cos^2\theta\,
e^{-i(\om_{\q,1}t-\q z)}\, \,\exp\left[-\frac{(z-v_{\q_1}t)^2}{4
S_1(t)} \right] \int_{-\q}^\infty \, d(\delta
k)\,\,\exp\left[-S_1(t)\left(\delta k -i \frac{(z-v_{\q,1}t)}{2
S_1(t)}\right)^2\right]
\\ \label{psi}
& + & A H\,\sin^2\theta\, e^{-i(\om_{\q,2}t-\q z)}\,
\,\exp\left[-\frac{(z-v_{\q_2}t)^2}{4 S_2(t)} \right]
\int_{-\q}^\infty \, d(\delta k)\,\,\exp\left[-S_2(t)\left(\delta k
-i \frac{(z-v_{\q_2}t)}{2 S_2(t)}\right)^2\right] \eea
with $S_i(t)\equiv \sigma_x^2 (1+i t/ T_i^{disp})$. Also,
$\sigma_x$ is a spread in the configuration space, defined by
$\sigma_x\sigma_{k}=\frac{1}{2}$ and $v_{\q,i}\equiv \q/\om_{\q,i}$.
To a good approximation we can  set $\int_{-\q}^\infty \approx
\int_{-\infty}^\infty$ and  obtain
\bea\non &&\int_{-\infty}^\infty d(\delta k)\,
e^{-S_i(t)\left(\delta k -i \frac{(z-v_{\q,1}t)}{2 S_i(t)}\ri)^2}\,
=\,\sqrt{\frac{4\pi\sigma^2_k}{1+ i t /T_i^{disp}}}\equiv I_i(t) \eea
Following the usual treatment \cite{Beuthe}, we now consider
propagation for $t\ll T_i^{disp}$. Then $I_i(t)$ becomes a
constant: $I= 2\sqrt{\pi}\sigma_{k}$. For propagation times beyond
$T_i^{disp}$, Laplace's method cannot be applied and one has to
resort to the method of stationary phase \cite{laplace,Beuthe}.

We are now in a position to calculate the interference term. We get, after some
rearrangements,
\bea\non \langle \nu_e |J_e^3(z,t)| \nu_e\rangle & \simeq
&|A|^2 H^2 |I|^2 \Big( \cos^4\theta\,
\exp\left[-\frac{(z-v_{\q,1}t)^2}{2\sigma_x^2}\right] +
\sin^4\theta\,
\exp\left[-\frac{(z-v_{\q,2}t)^2}{2\sigma_x^2}\right]\Big)
\\ \label{interf}
&+& 2 \,|A|^2 H^2|I|^2\,\Re\Big[\cos^2\theta\sin^2\theta\,
{e}^{-i\phi(t)-f(t)}\Big]\eea
where the phase $\phi_{12}$ and $f(t)$ are given by
\bea\label{phase}
\phi(t)\,=\,(\om_{\q,1}-\om_{\q,2})t\quad ; \quad
 f(t)\,=\,\frac{1}{4\sigma_x^2}
\left[(z-v_{\q,1}t)^2+(z-v_{\q,2}t)^2\right] \eea
We note that $\phi(t)$ and
$f(t)$ are the usual terms  obtained in the external wave packet
model in QFT \cite{Beuthe,giuntipak}.
We now proceed with time integration in order to finally get an expression
involving distance only.
\subsubsection{Time average}

When calculating the interference term, spatial oscillations results from the cross--term.
We show this explicitly by considering the first term in Eq.(\ref{interf}):
in the Laplace regime where $(t/T^{disp}_i)\approx 0$, if we assume that the experiment
runs for a very long time,
\bea\non
\int^\infty_{-\infty}\,
\exp\left[-\frac{(z-v_{\q,i}t)^2}{2\sigma_x^2}\right]\,dt\, =
\, \frac{\sqrt{2\pi}}{v_{\q,i}}\sigma_x
\eea
The first and second terms  of Eq.(\ref{interf})
are thus merely a collection of constants.

Let us now turn to the oscillation term. The dominant contributions in the time
integral come from $\exp[-f(t)]$, more precisely in the neighborhood of
the maximum value of $f(t)$, which occurs at
\bea
t_{max}=\left(\frac{v_{\q,1}+v_{\q,2}}{v_{\q,1}^2+v_{\q,2}^2}\right)z
\eea
Thus the integrand can be approximated as
\bea
{e}^{-i\phi(t)-f(t)} \, \simeq \,
\exp\left[-i\phi_{max}-f_{max}-i t \,\left.\frac{{d}\phi}{{d}t}
\right|_{t_{max}}- \frac{t^2}{2}\, \left.\frac{{d}^2
f}{{d}t^2}\right|_{t_{max}}\right]
\eea
\bea\label{coherence} f_{max}= f(t_{max})&=&\frac{z^2
\si_k^2(v_{\q,1}-v_{\q,2})^2}{v_{\q,1}^2+v_{\q,2}^2} \approx
z^2\left(\frac{\sqrt{2}\pi\sigma_{k}}{L^{osc}\q}\right)^2=\lf(\frac{z}{L^{coh}}\ri)^2
\\\label{osc} \phi_{max}=\phi(t_{max})&=&
\frac{z \q\left(v_{\q,2}^2-v_{\q,1}^2\right)}{v_{\q,1}v_{\q,2}
(v_{\q,1}^2+v_{\q,2}^2)}\approx - 2\pi\frac{z}{L^{osc}} \eea
to first order\footnote{We assume  that $v_{\q,i}\approx
1-\frac{m_i^2}{2\q^2}$. Thus  $v_{\q,1}^2+v_{\q,2}^2\approx 2 $,
$v_{\q,2}^2-v_{\q,1}^2 \approx - \triangle m^2/\q^2$ and
$v_{\q,1}v_{\q,2}\approx 1$.}. A coherence length
$L^{coh}=(L^{osc}\q)/ (\sqrt{2}\pi\sigma_{k})$ and an oscillation
length $L^{osc}=4\pi \q/\triangle m^2$  with $\De m^2=m_2^2 -m_1^2$,
having the same form as in
the standard approach, are thus recovered.  The other terms  can be
factorized as follows:
\bea
\frac{v_{\q,1}^2+v_{\q,2}^2}{4\sigma_x^2}
\left(t-i\frac{(\om_{\q,1}-\om_{\q,2})}{v_{\q,1}^2+v_{\q,2}^2}2\sigma_x^2\right)^2+
\sigma_x^2\frac{(\om_{\q,1}-\om_{\q,2})^2}{v_{\q,1}^2+v_{\q,2}^2}
\eea
The first term integrates to a constant: $\sqrt{2\pi}\sigma_x$. As for the second term,
it becomes $\frac{\sigma_x^2}{2}(\om_{\q,1}-\om_{\q,2})^2\simeq
2\pi^2\left(\frac{\sigma_x}{L^{osc}}\right)^2$
which gives the localization term\cite{Beuthe}.
Hence we can finally write down the (normalized)
highly-relativistic space oscillation formula
(compare to Ref.\cite{giuntipak}):
\bea\label{ours} \Phi_{\nu_e\to \nu_e}(z)&\simeq & 1-\frac{1}{2}\sin^2(2\theta)
\left\{1-\cos\left(2\pi\frac{z}{L^{osc}}\right)
\,\exp\lf[-\left(\frac{z}{L^{coh}}\right)^2-2\pi^2
\left(\frac{\sigma_x}{L^{osc}}\right)^2\ri]
\right\} \eea
Thus we have shown that the exact formula Eq.(\ref{flux1D}) reproduces the usual
oscillation formula in the relativistic limit.

\subsection{Numerical Calculations}

In this Section, we perform  numerical evaluations of
the exact oscillation formula Eq.(\ref{flux1D}) in order to compare it
with the approximate result Eq.(\ref{ours})
for sample values of the parameters.
We plot the two expressions in the maximal mixing case for a given $\si_k$ and for
different values of $\q$ and of the masses:
in the relativistic case (Fig.1) we observe a perfect agreement of the two formulas,
as expected.

We then explore the non-relativistic region and
observe that relevant deviations from the standard formula do indeed appear,
both in the oscillation amplitude and in the phase.
This is already evident in Figs.2,3 where we use $\q=100$ with $m_\1=1$, $m_\2=3$
in the first case and $m_\1=1$, $m_\2=10$ in the second case. The fact that
the two plots are simply scaled with respect to each other,
indicates that for these values of parameters the usual relation
coherence length vs. oscillation length:
$L^{coh}=(L^{osc}\q)/ (\sqrt{2}\pi\sigma_{k})$ is still valid with
$L^{osc}=4\pi \q/\triangle m^2$.
In Fig.4 we use $\q=50$ with $m_\1=1$, $m_\2=3$ and
as expected we observe a larger deviation
from the usual formula.

Note that the value of the Bogoliubov coefficient $|V_k|$ is very small in all the considered
cases\footnote{The values of $|V_\q|$ are: $1\times 10^{-3}$ (Fig.1),
$1\times 10^{-2}$ (Fig.2), $4\times 10^{-2}$ (Fig.3), $2\times 10^{-2}$ (Fig.4).},
so the observed corrections originate from  the ``standard'' terms
$\psi_\pm$  Eqs.(\ref{general1}),(\ref{general3}) in the oscillation formula
Eq.(\ref{flux1D}) rather than from the flavor vacuum contribution, which thus turns out to
be very difficult to detect when considering
(time-)averaged oscillation formulas like in the present case.

\vspace{0.5cm}

\centerline{\epsfysize=3.0truein\epsfbox{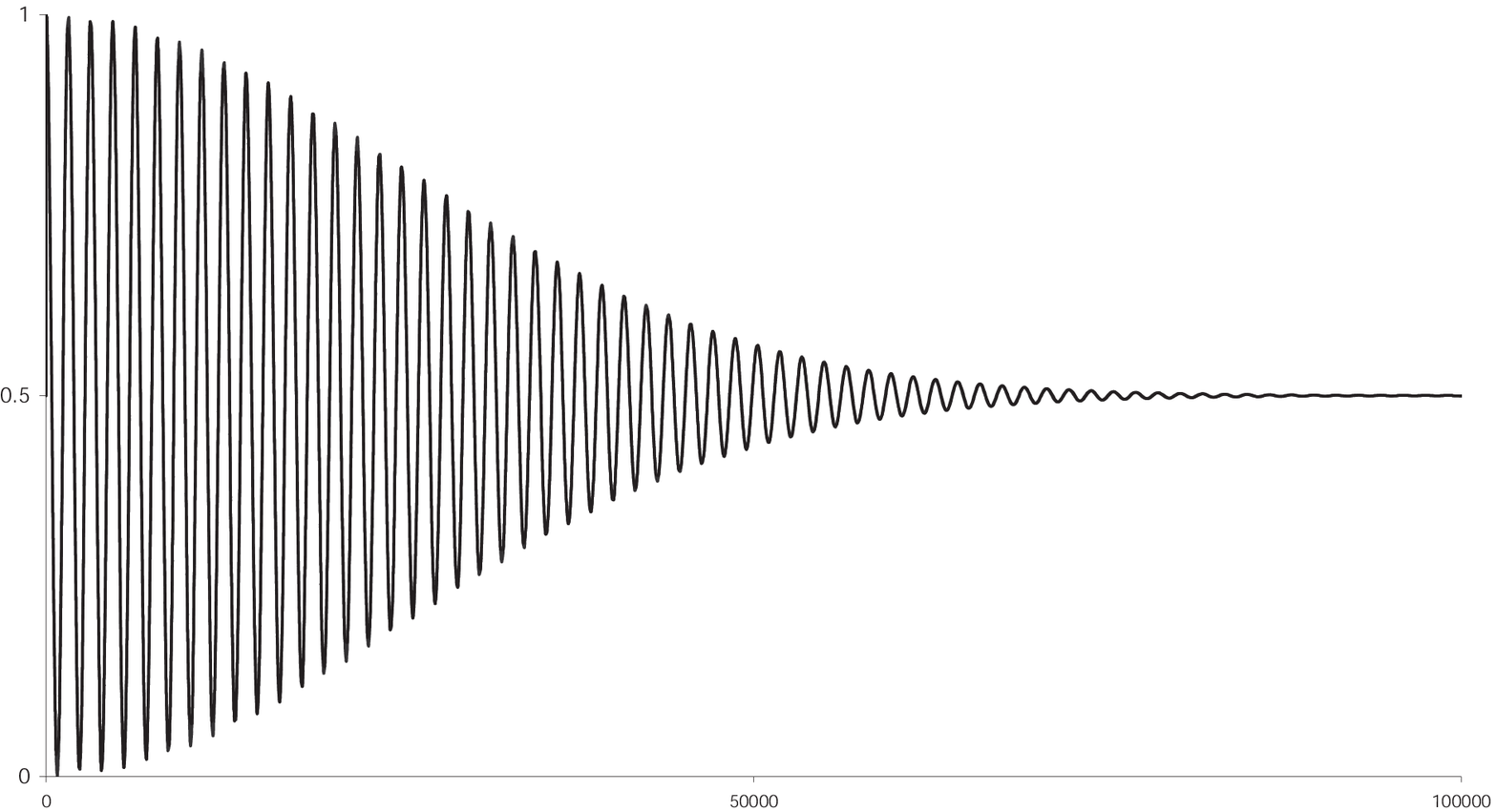}} \vspace{.2cm}
\begin{center}
{\small Figure 1:  Plot of the QFT flavor flux (solid line) against
the standard oscillation formula (dashed line) \\
for $\te=\pi/4$, $\si_k =10$, $m_\1=1$, $m_\2=3$, $\q=1000$.}
\end{center}

\vspace{.5cm}

\centerline{\epsfysize=3.0truein\epsfbox{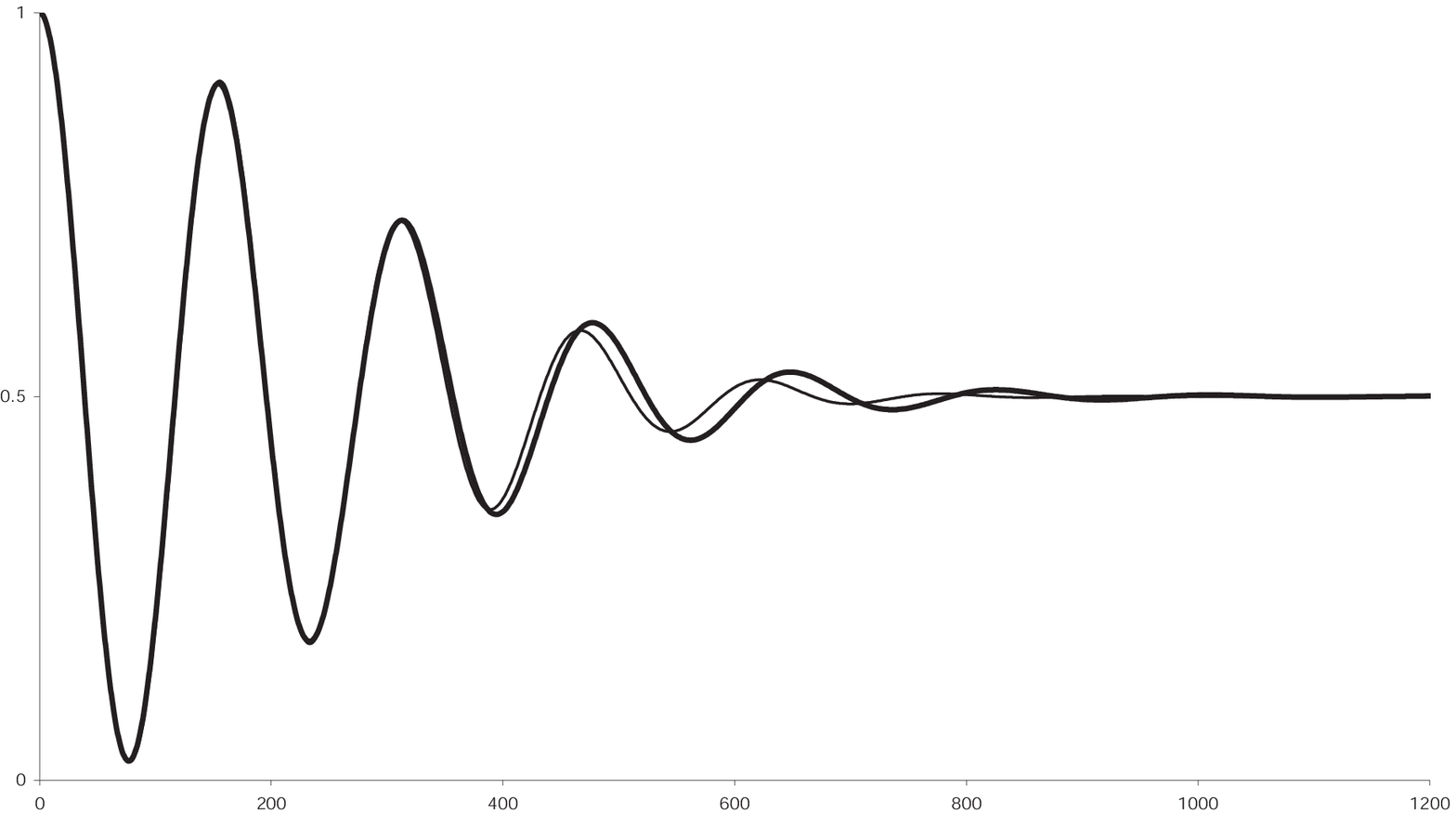}} \vspace{.2cm}
\begin{center}
{\small Figure 2:  Plot of the QFT flavor flux (thick line) against
the standard oscillation formula (light line) \\
for $\te=\pi/4$, $\si_k =10$, $m_\1=1$, $m_\2=3$, $\q=100$.}
\end{center}

\vspace{.5cm}

\centerline{\epsfysize=3.0truein\epsfbox{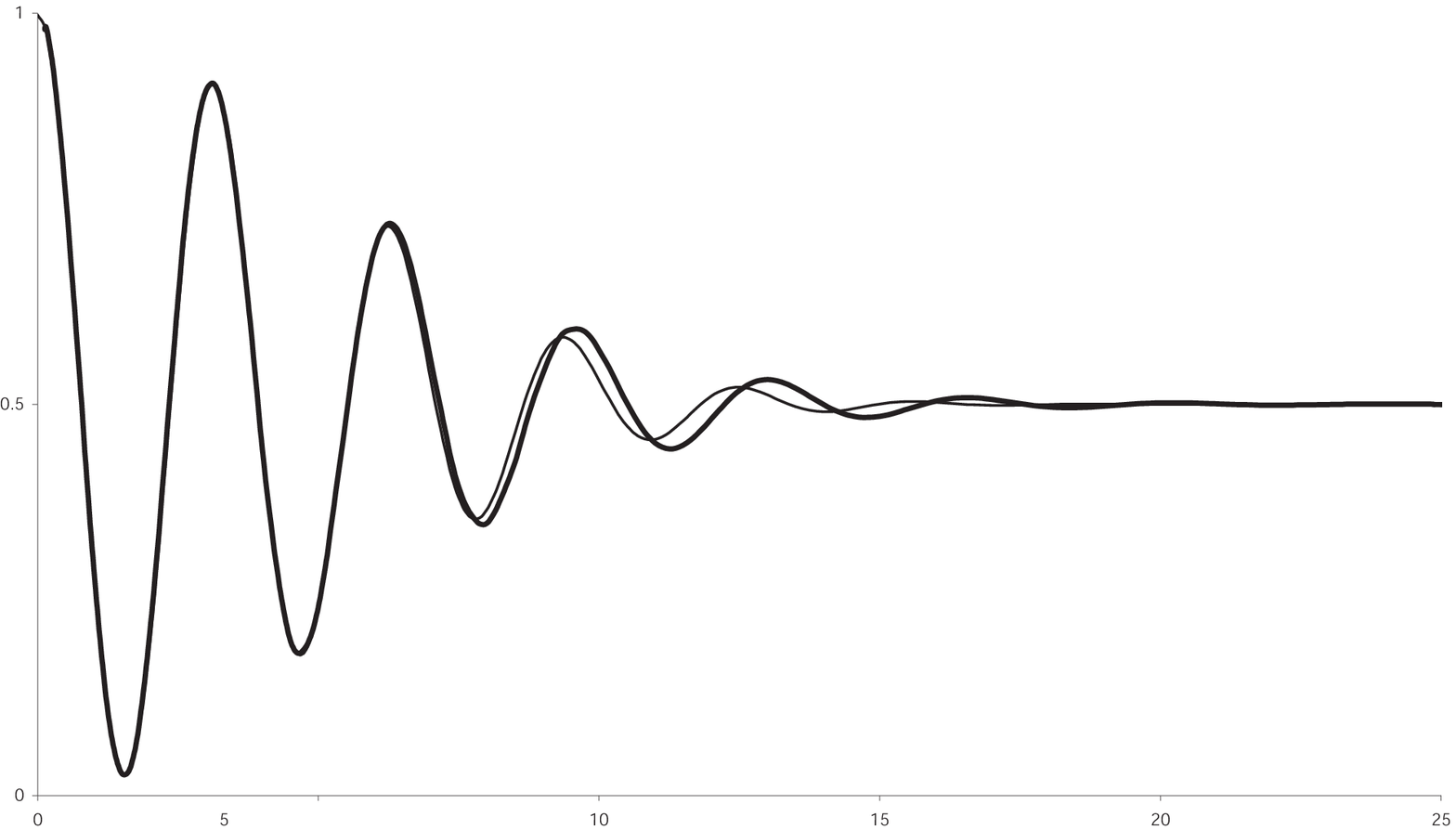}} \vspace{.2cm}
\begin{center}
{\small Figure 3:  Plot of the QFT flavor flux (thick line) against
the standard oscillation formula (light line) \\
for $\te=\pi/4$, $\si_k =10$, $m_\1=1$, $m_\2=10$, $\q=100$.}
\end{center}

\vspace{.5cm}

\centerline{\epsfysize=3.0truein\epsfbox{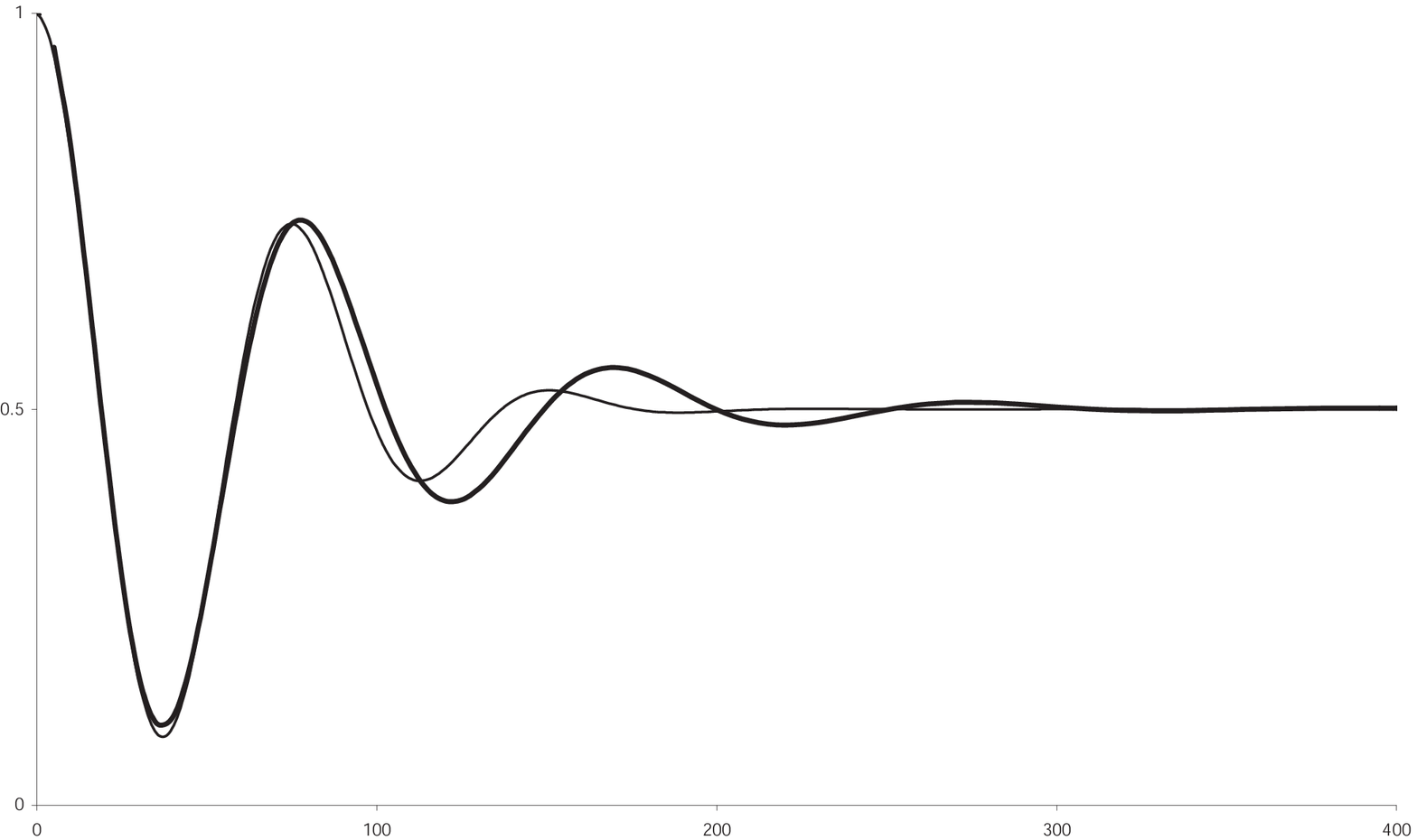}} \vspace{.2cm}
\begin{center}
{\small Figure 4:  Plot of the QFT flavor flux (thick line) against
the standard oscillation formula (light line) \\
for $\te=\pi/4$, $\si_k =10$, $m_\1=1$, $m_\2=3$, $\q=50$.}
\end{center}

A more complete discussion of these corrections and of their possible
phenomenological relevance will be given elsewhere \cite{BPT03}.

\section{Conclusions}

In this paper, we have for the first time derived a space-time dependent oscillation formula
directly from the relativistic currents for flavor fields. These currents have been recently
studied in Ref.\cite{currents}, a result which served as basis for the present analysis,
together with previous results on the Hilbert space for mixed fields\cite{BV95,BHV99}.

We first presented a general expression for the electron neutrino flux
in three dimensions
and then specialized to the case
with spherical symmetry, for which we were able to find a more
 explicit expression. In order to
perform further analysis and numerical evaluations,
we then considered the one-dimensional case with gaussian wave-packets,
which is also the one most frequently treated in literature.

Our formulation presents several advantages
with respect to existing treatments of neutrino oscillations
in Quantum Field Theory:
it is a very straightforward approach which is easy to relate to
practical experimental situations; it takes into account the non-trivial nature of the flavor
vacuum and flavor states are thus consistently defined. It takes explicitly into account the
full spin structure of neutrino states and does not resort to relativistic limit and/or
assumption of nearly degenerate masses for the energy eigenstates.

We have shown how, in different limits, our formula reproduce existing results. Thus,
in the case of relativistic neutrinos with nearly degenerate masses, we recover analytically
the standard space dependent expression for
neutrino oscillations \cite{giuntipak,Beuthe}, which is thus
once again confirmed from an independent approach.
Also, previous results on the flavor charges\cite{BHV99} have been recovered,
exhibiting the non-standard oscillation terms.

The numerical analysis shows that in the non-relativistic regime,
our formula  predicts significant deviations from the
standard oscillation formula \cite{giuntipak,Beuthe}, which on the other hand cannot
be expected to be valid in that region.
An analysis of the phenomenological implications of the
results obtained in this paper
will be presented elsewhere\cite{BPT03}.

\section{Acknowledgements}
M.B. thanks the ESF network COSLAB and  EPSRC for support. P.P.P.
and H.W.C.T. would like to thank the organizers of the
International  Workshop on the Quantum Field Theory of Particle
mixing and Oscillations, 13-15 June 2002, Vietri sul Mare (Italy)
and the Royal College of Science Association for financial
support. PPP also thanks the FCT-MCES and the ESF for their
financial support in the context of 3rd EC support framework.

\appendix

\section{Flavor Hilbert space}

The free fields $\nu_{1}(x)$ and $\nu_{2}(x)$ are written as ($t\equiv x_0$):
\bea
\lab{nui} \nu_{i}(x) = \sum_{r=1,2} \int
\frac{d^3\bk}{(2\pi)^\frac{3}{2}}  e^{i {\bf k}\cdot {\bf x}}
\lf[u^{r}_{{\bf k},i}(t) \al^{r}_{{\bf
k},i}\:+ v^{r}_{-{\bf k},i}(t) \bt^{r\dag }_{-{\bf k},i} \ri],
\qquad i=1,2 \;,
\eea
where $u^{r}_{{\bf k},i}(t)=e^{-i\om_{k,i} t}u^{r}_{{\bf k},i}$
and $v^{r}_{{\bf k},i}(t)=e^{i\om_{k,i} t}v^{r}_{{\bf k},i}$, with
$\om_{k,i}=\sqrt{|\bk|^2+m_i^2}$. The $\al^{r}_{{\bf k},i}$ and
the $\bt^{r }_{{\bf k},i}$, $i,r=1,2$ are the
annihilation operators for the vacuum state
$|0\ran_{1,2}\equiv|0\ran_{1}\otimes \,|0\ran_{2}$: $\al^{r}_{{\bf
k},i}|0\ran_{12}= \bt^{r }_{{\bf k},i}|0\ran_{12}=0$.
 The
anticommutation relations are:
\bea\lab{2.3}
&&\{\nu^{\al}_{i}({\bf
x}), \nu^{\bt\dag }_{j}({\bf y})\}_{t=t'} =  \,
\de^{3}({\bf{x}}-{\bf{y}}) \de_{\al\bt} \de_{ij} \;, \qquad
\al,\bt=1,..,4 \;,
\\ \lab{2.4}
&& \{\al^{r}_{{\bf k},i}, \al^{s\dag }_{{\bf p},j}\} =
\de^3 ({\bk - \bp})\de _{rs}\de_{ij} \,  ;\qquad \{\bt^{r}_{{\bf
k},i}, \bt^{s\dag}_{{\bf p},j}\} = \de^3({\bk - \bp})
\de_{rs}\de_{ij},\;\;\;\; i,j=1,2\;. \eea
All other anticommutators are zero. The orthonormality and
completeness relations are:
\bea && u^{r\dag}_{{\bf k},i} u^{s}_{{\bf k},i} = v^{r\dag}_{{\bf
k},i} v^{s}_{{\bf k},i} = \de_{rs} \;,\qquad  u^{r\dag}_{{\bf
k},i} v^{s}_{-{\bf k},i} = v^{r\dag}_{-{\bf k},i} u^{s}_{{\bf
k},i} = 0\;, \quad ;\quad \sum_{r}(u^{r}_{{\bf
k},i}u^{r\dag}_{{\bf k},i}  + v^{r}_{-{\bf k},i}v^{r\dag}_{-{\bf
k},i} ) = \ide_2 \;.
\eea
where $\ide_n$ is the $n \times n$ unit matrix.
Eq.(\ref{fermix}) can be recast in the form:
\bea
\nu_\si(x) & \equiv & G^{-1}_\te(t) \, \nu_i(x) \, G_\te(t) \, ,
\\ [2mm]
G_{\te}(t)& =&
exp\lf[\te \int d^{3}{\bf x}  \lf(\nu_{1}^{\dag}(x) \nu_{2}(x) -
\nu_{2}^{\dag}(x) \nu_{1}(x) \ri)\ri]\;,
\eea
with $(\si,i)=(e,1) , (\mu,2)$.
The generator $G_\te(t)$ does not leave invariant
the vacuum $|0 \ran_{1,2}$:
\bea
\lab{2.22} |0 (t)\ran_{e,\mu} = G^{-1}_\te(t)\; |0
\ran_{1,2}\;.
\eea
We will refer to $|0(t) \ran_{e,\mu}$ as the flavor vacuum: it is
orthogonal to $|0 \ran_{1,2}$ in the infinite volume limit\cite{BV95}.
We define the flavor annihilators, relative to
the fields $\nu_{e}(x)$ and $\nu_{\mu}(x)$ as\footnote{The
annihilation of the flavor vacuum at each time is expressed as:
$\al^{r}_{{\bf k},e}(t)|0(t) \ran_{e,\mu}=G^{-1}_\te(t) \,
\al^{r}_{{\bf k},1}|0 \ran_{1,2}=0$. }
\bea
\al^{r}_{{\bf k},\si}(t) \equiv G^{-1}_{\te}(t)\al^{r}_{{\bf k},i}(t)
G_{\te}(t)\quad ;\quad \bt^{r\dag}_{{-\bf k},\si}(t)\equiv
 G^{-1}_{\te}(t)  \bt^{r\dag}_{{-\bf k},i}(t)
G_{\te}(t)
\eea
with $(\si,i)=(e,1) , (\mu,2)$.
The flavor fields can be expanded as:
\bea \label{flavfields} \nu_\si(x)&=& \sum_{r=1,2} \int \frac{d^3
\bk}{(2\pi)^\frac{3}{2}} \lf[ u^{r}_{{\bf k},i}(t) \al^{r}_{{\bf k},\si}(t)
+    v^{r}_{-{\bf k},i}(t) \bt^{r\dag}_{-{\bf k},\si}(t) \ri]
e^{i {\bf k}\cdot{\bf x}}\,. \eea
with $(\si,i)=(e,1) , (\mu,2)$.
The flavor annihilation operators are defined as
operators
\bea
\al^{r}_{{\bf k},e}(t)&=&\cos\te\;\al^{r}_{{\bf
k},1}\;+\;\sin\te\;\sum_{s}\lf[ u^{r\dag}_{{\bf k},1}(t)
u^{s}_{{\bf k},2}(t)\; \al^{s}_{{\bf k},2}\;+\; u^{r\dag}_{{\bf
k},1}(t) v^{s}_{-{\bf k},2}(t)\;
\bt^{s\dag}_{-{\bf k},2}\ri]\\
\lab{2.35}
\al^{r}_{{\bf k},\mu}(t)&=&\cos\te\;\al^{r}_{{\bf
k},2}\;-\;\sin\te\;\sum_{s}\lf[ u^{r\dag}_{{\bf k},2}(t)
u^{s}_{{\bf k},1}(t)\; \al^{s}_{{\bf k},1}\;+\; u^{r\dag}_{{\bf
k},2}(t) v^{s}_{-{\bf k},1}(t)\;
\bt^{s\dag}_{-{\bf k},1}\ri]\\
\bt^{r}_{-{\bf k},e}(t)&=&\cos\te\;\bt^{r}_{-{\bf
k},1}\;+\;\sin\te\;\sum_{s}\lf[ v^{s\dag}_{-{\bf k},2}(t)
v^{r}_{-{\bf k},1}(t)\; \bt^{s}_{-{\bf k},2}\;+\; u^{s\dag}_{{\bf
k},2}(t) v^{r}_{-{\bf k},1}(t)\; \al^{s\dag}_{{\bf
k},2}\ri]\\
\bt^{r}_{-{\bf k},\mu}(t)&=&\cos\te\;\bt^{r}_{-{\bf
k},2}\;-\;\sin\te\;\sum_{s}\lf[ v^{s\dag}_{-{\bf k},1}(t)
v^{r}_{-{\bf k},2}(t)\; \bt^{s}_{-{\bf k},1}\;+\; u^{s\dag}_{{\bf
k},1}(t) v^{r}_{-{\bf k},2}(t)\; \al^{s\dag}_{{\bf k},1}\ri] \eea
In the reference frame where
${\bf{k}}$ is collinear with ${\hat {\bf k}} \equiv (0,0,1)$, the spins decouple and we
have:
\bea
\al^{r}_{{\bf k},e}(t)&=&\cos\te\;\al^{r}_{{\bf
k},1}\;+\;\sin\te\;\lf( U_{{\bf k}}^{*}(t)\; \al^{r}_{{\bf
k},2}\;+\;\ep^{r}_{\bf k}\;
V_{{\bf k}}(t)\; \bt^{r\dag}_{-{\bf k},2}\ri) \\
\lab{2.36}
\al^{r}_{{\bf k},\mu}(t)&=&\cos\te\;\al^{r}_{{\bf
k},2}\;-\;\sin\te\;\lf( U_{{\bf k}}(t)\; \al^{r}_{{\bf
k},1}\;-\;\ep^{r}_{\bf k}\;
V_{{\bf k}}(t)\; \bt^{r\dag}_{-{\bf k},1}\ri)  \\
\bt^{r}_{-{\bf k},e}(t)&=&\cos\te\;\bt^{r}_{-{\bf
k},1}\;+\;\sin\te\;\lf( U_{{\bf k}}^{*}(t)\; \bt^{r}_{-{\bf
k},2}\;-\;\ep^{r}_{\bf k}\;
V_{{\bf k}}(t)\; \al^{r\dag}_{{\bf k},2}\ri)
\\
\bt^{r}_{-{\bf k},\mu}(t)&=&\cos\te\;\bt^{r}_{-{\bf
k},2}\;-\;\sin\te\;\lf( U_{{\bf k}}(t)\; \bt^{r}_{-{\bf
k},1}\;+\;\ep^{r}_{\bf k}\; V_{{\bf k}}(t)\; \al^{r\dag}_{{\bf k},1}\ri)
\eea
where $\ep^{r}_{\bf k}\equiv (-1)^{r + {\bf k}\cdot {\hat {\bf k}} + 1}$  and
$U_{{\bf k}}(t)$, $V_{{\bf k}}(t)$ are Bogoliubov coefficients given by:
\bea
&&U_{{\bf k}}(t)\equiv u^{r\dag}_{{\bf k},2}(t)u^{r}_{{\bf
k},1}(t)=
v^{r\dag}_{-{\bf k},1}(t)v^{r}_{-{\bf k},2}(t)\,=\,|U_{{\bf k}}|\;e^{i(\om_{k,2}-\om_{k,1})t}
\\ [2mm]\lab{2.37}
&&V_{{\bf k}}(t)\equiv \ep^{r}_{\bf k}\;u^{r\dag}_{{\bf
k},1}(t)v^{r}_{-{\bf k},2}(t)= -\ep^{r}_{\bf k}\;u^{r\dag}_{{\bf
k},2}(t)v^{r}_{-{\bf k},1}(t)\,=\,|V_{{\bf k}}|
\;e^{i(\om_{k,2}+\om_{k,1})t}
\\ [2mm]
&&|U_{{\bf k}}|=\lf(\frac{\om_{k,1}+m_{1}}{2\om_{k,1}}\ri)^{\frac{1}{2}}
\lf(\frac{\om_{k,2}+m_{2}}{2\om_{k,2}}\ri)^{\frac{1}{2}}
\lf(1+\frac{|{\bf{k}}|^{2}}{(\om_{k,1}+m_{1})(\om_{k,2}+m_{2})}\ri)
\\
\lab{2.39} &&
|V_{{\bf k}}|=\lf(\frac{\om_{k,1}+m_{1}}{2\om_{k,1}}
\ri)^{\frac{1}{2}}
\lf(\frac{\om_{k,2}+m_{2}}{2\om_{k,2}}\ri)^{\frac{1}{2}}
\lf(\frac{{|{\bf k}|}}{(\om_{k,2}+m_{2})}-
\frac{{|{\bf k}|}}{(\om_{k,1}+m_{1})}\ri)
\eea
satisfying $|U_{{\bf k}}|^{2}+|V_{{\bf k}}|^{2}=1$.

\section{Chiral representation and useful relations}
\label{chiral}
\bea
u_{k,i}^{r} =\chi\,_{i}
\left(\begin{array}{c}
 ( 1 + \frac{\vec{\sigma}\cdot{\bf k}}{\omega_{k,i}+m_{i}}) \xi^{r} \\ \\
( 1 - \frac{\vec{\sigma}\cdot\bk}{\omega_{k,i}+m_{i}}) \xi^{r}
\end{array}\right)\;\;\;,\;\;\;
v_{k,i}^{r} =\chi\,_{i} \left(\begin{array}{c}
(1 + \frac{\vec{\sigma}\cdot\bk}{\omega_{k,i}+m_{i}}) \xi^{r} \\ \\
( -1 + \frac{\vec{\sigma}\cdot\bk}{\omega_{k,i}+m_{i}}) \xi^{r}
\end{array}\right) \;,\quad r=1,2
\\ [2mm]
 \xi^{1}=\left(\begin{array}{c} 1 \\ 0
\end{array}\right)\;\;,\;\; \xi^{2}=\left(\begin{array}{c} 0 \\ 1
\end{array}\right)\;\;\;,\;\;\; \chi\,_{i}\equiv
\left(\frac{\omega_{k,i}+m_{i}}{4\omega_{k,i}}\right)^{\frac{1}{2}}
\;\;\;,\;\;\; i=1,2.
\eea
\bea
&&v_{-k,1}^{1\dag} u_{k,2}^{1}=
- v_{-k,1}^{2\dag} u_{k,2}^{2}= 2 \chi\,_{1}\chi\,_{2}
\left(\frac{-k_{3}}{\omega_{k,1}+m_{1}}
+\frac{k_{3}}{\omega_{k,2}+m_{2}} \right)
\\
&&v_{-k,1}^{1\dag} u_{k,2}^{2}=
\left(v_{-k,1}^{2\dag} u_{k,2}^{1}\right)^{*}
= 2\chi\,_{1}\chi\,_{2} \left(\frac{-k_-}{\omega_{k,1}+m_{1}}
+\frac{k_-}{\omega_{k,2}+m_{2}}\right)
\eea
and $u_{k,i}^{r\dag} u_{k,j}^{s}= 0$ for $r \neq s$.
We define:
\bea \Ga^0 = \ide_4 \quad , \quad
\Ga^i=\left(\begin{array}{cc} \sigma^i&0\\0& -\sigma^i
\end{array}\right)\,, \quad i = 1,2,3.
\eea
Notice that $\Ga^1$ and $\Ga^2$ can be decomposed as follows:
\bea \label{sigma1}
\Gamma^i=
\sum\limits_{j=1}^4\,\eta_j\eta_j^{\dag}\, - \, \la_i \la_i^{\dag}
\, - \, \vartheta_i \vartheta_i^{\dag}
\eea
with
\bea
\la_1=\lf(\begin{array}{c} 1\\ -1 \\ 0\\ 0\end{array}\ri) \quad, \quad
\vartheta_1=\lf(\begin{array}{c} 0\\ 0 \\ 1\\ 1\end{array}\ri)
\quad, \quad
\la_2=\lf(\begin{array}{c} 1\\ -i \\ 0\\ 0\end{array}\ri) \quad, \quad
\vartheta_2=\lf(\begin{array}{c} 0\\ 0 \\ 1\\ i\end{array}\ri)
\eea

\section{Vacuum expectation value of the current}

\label{appenvac}

We show here that  $_{e,\mu}\langle0|J_e^{\mu}(\bx,t)|0\rangle_{e,\mu}=0$.
Let us consider for example the following quantity:
\bea\label{example}
&&\,_{e,\mu}\langle0| \,
\alpha_{\bp,e}^{r\dag}(t)\alpha_{\bk,e}^s(t)\,|0\rangle_{e,\mu}
\,=\;_{1,2}\langle0|\:G_\te(0)\,
\alpha_{\bp,e}^{r\dag}(t)\alpha_{\bk,e}^s(t)
G^{-1}_\te(0)\;
|0 \ran_{1,2}\;
\eea
Now let us define $ {\ti \alpha}_{\bk,e}^{r}(t) \, \equiv\,
G_\te(0) \, \alpha_{\bk,e}^r(t) \, G^{-1}_\te(0)$ and h.c. Of course,
$ {\ti \alpha}_{\bk,e}^{r}(0) \, =\,
\alpha_{\bk,1}^r$. It is easy to realize that (see Eqs.(\ref{2.35})):
\bea
\lf\{{\ti \alpha}_{\bk,e}^{r}(t)\,,\,{\ti \alpha}_{\bp,e}^{s\dag}(t')\ri\}
\, =\, 0 \qquad for \quad \bk \neq \bp , \quad \forall \; t, t',r, s.
\eea
This in turn implies that the quantity (\ref{example}) vanishes for
$\bk \neq \bp$. Since this is valid for any two flavor operators,
we only need to calculate the various terms
for the (flavor) vev of the current  for equal momenta. We thus write:
\bea
\,_{e,\mu}\langle 0|J_e^{\mu}(\bx,t)|0\rangle_{e,\mu}=
\int d^3\bk \,  _{e,\mu}\langle0|J_e^{\mu}(\bk,t)|0\rangle_{e,\mu}=0.
\eea
The flavor vacuum is invariant under rotations, so we can choose without
loss of generality the most
suitable spatial reference frame for each of the terms in the
momentum integration.
In our case, this is the one  for which
${\bf{k}}$ is collinear to $(0,0,1)$. In this frame, we obtain the relations:
\bea \non
&&\,_{e,\mu}\langle0|\, \al^{r\dag}_{{\bf k},e}(t)\al^{s}_{{\bf k},e}(t)\,|0\rangle_{e,\mu}
\,=\,_{e,\mu}\langle0|\, \bt^{r\dag}_{-{\bf k},e}(t)\bt^{s}_{-{\bf k},e}(t)\,|0\rangle_{e,\mu}
 \\ \label{expect1}
&&= \, \de^{rs}\,4 |V_{{\bf k}}|^{2}\, \sin^2\te\,
\lf[\cos^2\te \,\sin^2\lf(\frac{\om_1+\om_2}{2}t\ri)
+ \sin^2\te |U_{{\bf k}}|^{2}\, \sin^2\lf(\om_2 t\ri)\ri]
\\ [2mm]\non
&&\,_{e,\mu}\langle0|\, \bt^{r}_{-{\bf k},e}(t)\al^{s}_{{\bf k},e}(t)\,|0\rangle_{e,\mu}
=\lf(\,_{e,\mu}\langle0|\, \al^{r\dag}_{{\bf k},e}(t)
\bt^{r\dag}_{-{\bf s},e}(t)\,|0\rangle_{e,\mu}\ri)^*
 \\ \label{expect2}
& &= \, \de^{rs}\, \ep^r_{\bf k} \sin^2\te\, e^{2 i \om_1 t}
\,|U_{{\bf k}}| \,  |V_{{\bf k}}| \,
\lf[\lf(1+ e^{-2 i \om_1 t} - 2 e^{-2 i (\om_1 +\om_2) t}\ri)\cos^2\te \,
+ 2 i \sin\lf(\om_2 t\ri)\lf(e^{- i \om_2 t}|U_{{\bf k}}|^2+
e^{i \om_2 t}|V_{{\bf k}}|^2\ri)\sin^2\te  \, \ri]
\eea
Because of relation (\ref{expect1}) and the orthonormality
relations for the spinors, we easily realize
that $_{e,\mu}\langle 0|J_e^{0}(\bk,t)|0\rangle_{e,\mu}=0$.

As for  $\,_{e,\mu}\langle 0|J_e^{3}(\bk,t)|0\rangle_{e,\mu}$,
we need to consider:
\bea\label{spin1}
&& u_{{\bf k},i}^{r\dag}(t) \Ga^3 u_{{\bf k},i}^{s}(t)\,=
\, - v_{-{\bf k},i}^{r\dag}(t) \Ga^3 v_{-{\bf k},i}^{s}(t)\,=\,
\de^{rs}\,\frac{k_3}{\om_{k,i}}
\\ \label{spin2}
&& u_{{\bf k},i}^{r\dag}(t) \Ga^3 v_{-{\bf k},i}^{r}(t)\,=
\,\lf(v_{-{\bf k},i}^{r\dag}(t) \Ga^3 u_{{\bf k},i}^{r}(t)\ri)^*\,=\,
- (-1)^r \,\de^{rs}\,\frac{m_i}{\om_{k,i}} \,e^{2 i \om_{k,i}t}
\eea
As a consequence of relations (\ref{expect1}), (\ref{expect2}) and
(\ref{spin1}), (\ref{spin2}), the expectation
value $_{e,\mu}\langle0|J_e^{3}(\bk,t)|0\rangle_{e,\mu}$
is an odd function of ${\bf k}$ and
therefore its integral vanishes.

Note that $_{e,\mu}\langle0|J_e^{1,2}(\bk,t)|0\rangle_{e,\mu}$ vanish identically in the
chosen reference frame, because of:
\bea
&& u_{{\bf k},i}^{r\dag}(t) \Ga^j u_{{\bf k},i}^{s}(t)\,=
\, - v_{-{\bf k},i}^{r\dag}(t) \Ga^j v_{-{\bf k},i}^{s}(t)\,=\,
u_{{\bf k},i}^{r\dag}(t) \Ga^j v_{-{\bf k},i}^{s}(t)\,=
\,0 \quad , \quad j=1,2.
\eea

\section{Explicit Forms}
\label{explicit}

\bea
&&{\cal A}^r_j({\bx},t) \,=\, A\int\frac{d^3
\bk}{(2\pi)^\frac{3}{2}}\,\,e^{i\bk.\bx}f(\bk) \sum_s\,\lf[\eta_j^{\dag}
u_{\bk,1}^s(t)\ri]\, \de^{s r} X_{\bk,e}(t)
\\[2mm]
&&{\cal B}^r_j({\bx},t) \,= \, A\int\frac{d^3
\bk}{(2\pi)^\frac{3}{2}}\,\,e^{i\bk \cdot\bx}f(\bk)\sum_s\,\lf[\eta_j^{\dag}
v_{-\bk,1}^s(t)\ri]\,(\si\cdot\bk)^{s r} Y_{\bk,e}(t)
\eea
\bea\non {\cal A}^1_1 &=&A\int\frac{d^3 \bk}{(2\pi)^\frac{3}{2}}
\,e^{i\bk \cdot\bx}\,f(\bk)\, \chi_1 \left(1+k_3/\Om_1\right)X_{\bk,e}(t)
\quad ; \quad {\cal B}^1_1 = A\int\frac{d^3 \bk}{(2\pi)^\frac{3}{2}}
\,e^{i\bk \cdot\bx}\,f(\bk)\,\chi_1\left(k_3- |\bk|^2/\Om_1 \right)
Y_{\bk,e}(t)
\\ \non
{\cal A}^1_2 &=& - {\cal A}^1_3 \,=\, A\int\frac{d^3
\bk}{(2\pi)^\frac{3}{2}}
\,e^{i\bk \cdot\bx}\,f(\bk)\,\chi_1\left(k_+/\Om_1\right)X_{\bk,e}(t) \quad ;
\quad {\cal B}^1_2 = -{\cal B}^1_3 = A\int\frac{d^3 \bk}{(2\pi)^\frac{3}{2}}
\,e^{i\bk \cdot\bx}\,f(\bk)\,\chi_1\ k_+Y_{\bk,e}(t)
\\ \label{4terms}
{\cal A}^1_4 &=&A\int\frac{d^3 \bk}{(2\pi)^\frac{3}{2}}
\,e^{i\bk \cdot\bx}\,f(\bk)\,\chi_1\left(1-k_3/\Om_1\right)X_{\bk,e}(t) \quad
; \quad {\cal B}^1_4 = -A\int\frac{d^3 \bk}{(2\pi)^\frac{3}{2}}
\,e^{i\bk \cdot\bx}\,f(\bk)\,\chi_1\left(|\bk|^2/\Om_1 + k_3
\right)Y_{\bk,e}(t) \eea
where $\Om_1 \equiv \om_{k,1}+m_1 $ and $k_+ = k_1 + i k_2$ , $k_-
= k_1 - i k_2$.

For ${\bf{k}}$  collinear with ${\hat {\bf k}} \equiv (0,0,1)$ or
in the one dimensional case, we have
\bea
\{\bt_{-\bk,e}^{s\dag}(t),\,\akd\} = \,\de^{sr}\ep^{r}_{\bf k}
|U_{\bk}||V_{\bk}|\sin^2\theta
\left[e^{i\om_{k,2}t}-e^{-i\om_{k,2}t}\right]e^{-i\om_{k,1}t}\,.
\eea

\end{document}